\documentclass{aa}     
\usepackage{graphicx}  

\usepackage{times}

\def\sectionApp#1{ \refstepcounter{section}\section*{Appendix: #1}}
\def\LabelFig#1#2{ \refstepcounter{figure}\count100=\thefigure		
	           \def\thefigure{\the\count100 #1}\label{#2}	
                   \addtocounter{figure}{-1}}				

\begin{document}

\thesaurus{09(06.18.2;03.13.4;06.09.1;06.15.1)}

\title{Non linear regularization for helioseismic inversions}

\subtitle{ Application for the study of the solar tachocline}

\author{T. Corbard\inst{1,2} \and L. Blanc-F\'eraud\inst{3} \and G. Berthomieu\inst{1} \and J. Provost\inst{1}}

\institute{Laboratoire G.-D. Cassini, CNRS UMR 6529, Observatoire de la C\^ote 
d'Azur, BP 4229, 06304  Nice Cedex 4, France \and 
NCAR/High Altitude Observatory, PO Box 3000, Boulder, CO 80307, 
USA (present address)\and 
Projet Ariana, CNRS/INRIA/UNSA, 2004 route des Lucioles, BP 93, 
     06902 Sophia Antipolis Cedex, France}

\offprints{T. Corbard}

\mail{corbard@hao.ucar.edu}

\date{Received 31 August 1998 /  Accepted 7 January 1999}

\maketitle

\begin{abstract}
%
Inversions of rotational splittings have shown that there exists at the 
base of the solar convection zone a region called the tachocline in which 
high radial gradients of the rotation rate occur.
The usual linear regularization methods tend to smooth out any high gradients 
in the solution, and may not be appropriate for the study of this zone.
In this paper we use, in the helioseismic context of rotation inversions,
regularization methods that have been developed for edge-preserving 
regularization in computed imaging.
It is shown from Monte-Carlo simulations that this approach can lead directly 
to results similar to those reached by linear inversions which however required
some assumptions on the shape of the transition in order to be deconvolved.
The application of this method to LOWL data leads to a very thin tachocline. 
From the  discussions on  the parameters entering the inversion and the
Monte-Carlo simulations, our conclusion is that the tachocline width is very
likely below $0.05R_\odot$ which lowers our previous estimate of
$0.05\pm 0.03R_\odot$ obtained from the same  dataset (Corbard et al. 1998).

\keywords{ Sun: rotation -- Methods: numerical  --  Sun: interior -- 
Sun: oscillations }

\end{abstract}

\section{Introduction}
%
Helioseismic inverse problems consist in using the properties (namely 
frequencies or frequency splittings) of the oscillation pattern observed at 
the surface of the Sun in order to infer the internal variation of solar 
physical properties like the rotation rate, the 
sound speed or the density (see e.g. Gough \& Thompson (\cite{gough91}) for 
a review).
These problems can be expressed in terms of  integral equations which 
represent  ill-posed problems in the sense of Hadamard (\cite{hadamard23}).

 The traditional approach used to override this difficulty consists in
\emph{regularizing the problem} by
adding some a priori information on the solution 
(e.g. Kirsch \cite{kirsch}; Craig \& Brown \cite{craig}).
 The well known Tikhonov regularization method 
(Phillips \cite{phillips62}; Twomey \cite{towney63}; Tikhonov 
\cite{tikhonov63}) 
assumes a global smoothness of the solution by minimizing the norm
of its  derivative at a given order.  
Nevertheless,
 inversions for the solar rotation  have shown 
(see e.g. Thompson et~al. (\cite{thompson:gong96}) and 
Schou et~al. (\cite{schou:mdi98})  for the latest results) 
that  high gradients exist in the  solar rotation profile near
the surface  and at the base of the convection zone 
in the so-called \emph{solar tachocline} (Spiegel \& Zahn \cite{spiegel92}).

Therefore, enforcing  global smoothness  a priori may
not be appropriate for the study of these zones, which
are  of particular interest for the study
of solar dynamics. As a matter of fact, the tachocline
represents a thin zone where the differential rotation
of the convection zone becomes rigid in the radiative
interior. It is thought to be the place where the
solar dynamo originates and its precise structure is
an important test for angular momentum transport theories.
 More precisely, the thickness 
of the tachocline can be related to the horizontal component
of the turbulent viscosity and may be used  as an important 
observational constraint on the dynamics  properties of 
the turbulence (Spiegel \& Zahn \cite{spiegel92}; Gough \& Sekii 
\cite{gough_sekii}; Elliot \cite{elliott}).  

Several works have already 
been performed to infer the fine structure of the tachocline
(Kosovichev \cite{kosovichev96},\cite{kosovichev98};  Basu \cite{basu97}; 
  Charbonneau et al. \cite{charbonneau98}; 
Antia et al. \cite{antia98}; Corbard et al. \cite{corbard98}) 
using both forward analysis and inverse 
techniques.
For the inverse approach,
 it may be interesting to change the  global  constraint
which tends to smooth out every high gradients in the 
solution and to find a way to preserve these gradients
 in the inversion process. 
A first attempt in that direction has been carried out 
(Corbard et al. \cite{corbard98}) by
using a nonlinear regularization term based on the  Piecewise 
Polynomials-Truncated Singular Value Decomposition (PP-TSVD)
method (Hansen \& Mosegaard \cite{PPTSVD}) which uses  a $1$-norm.
 Nevertheless
it has been shown that this method can produce
solutions with sharp discontinuities even if the 
width of the tachocline  is relatively large. This leads
to very large error bars on the inferred width of
 high gradient zones and complicates the interpretation 
of the results obtained from real observations. 
More elaborate nonlinear techniques have been developed 
for edge-preserving regularization
in computed imaging.
 In this paper, we investigate their use in the helioseismic context.

 Section~\ref{sec:pb} briefly recalls how the
solar internal rotation can be related to the frequency splittings
determined from  helioseismic measurements, and introduces 
the corresponding discretized inverse problem. The non linear approach
of regularization in inverse techniques is introduced in 
Sect.~\ref{sec:Reg_non_lin}, and the computational aspects  
are discussed. For the particular case of  solar
rotation inversion and the determination of the tachocline width,  
Sects.~\ref{sec:choix_reg_param} and \ref{sec:mc} 
present how the regularizing parameters 
have  been chosen and give the results of Monte-Carlo simulations for
the estimation of the uncertainty on the tachocline width. Finally,
 the results obtained with LOWL data are discussed
in Sect.~\ref{sec:real_data}, and we conclude in Sect.~\ref{sec:conclu}.  

\section{The astrophysical problem and its discretization}\label{sec:pb}
%
The internal rotation $\Omega(r,\theta)$ of the Sun expressed as a function of 
the solar radius $r$
and colatitude $\theta$ can be related (Hansen et al. \cite{hansen77}), 
through a 2D integral equation,
to the observed frequency splittings
$\Delta\nu_{nlm}$
where  each mode of  solar acoustic oscillations is characterized by 
 the degree 
$l$, the radial order $n$ and the azimuthal order $m$ ($-l\le m\le l$).

In this work, we focus on the application of non linear inversion to the 
description of the tachocline profile in the equatorial plane.
 The so-called `tachocline parameters' 
 are obtained by fitting,  between $0.4R_\odot$
and $0.8R_\odot$, the solution  $\bar\Omega(r)$
of the inversion 
by an error function ($erf$) of the form:
\begin{equation}\label{eq:fiterf}
\Omega_{fit}(r)\!=\!\bar\Omega_0\!+\!{\bar\Omega_1-\bar\Omega_0\over 2}
 \left(\!1\!+\! erf\!\left(\! {r-\bar r_c\over 0.5 \bar w}\!\right)\right)
\!+\!\bar\alpha(r-0.7).
\end{equation}
This defines five tachocline parameters: $\bar\Omega_0$, $\bar\Omega_1$,
$\bar r_c$, $\bar w$ and $\bar\alpha$. The coefficient $\bar\alpha$ has been 
introduced, following Antia et al. (\cite{antia98}), in order
to take into account the linear behaviour sometimes 
found for the rotation rate in the convection zone just above the transition
in the equatorial plane (see Sect.~\ref{sec:real_data}).

Thus, we consider the 1D problem of inferring
the solar equatorial rotation profile $\Omega_{eq}=\Omega(r,90^\circ)$
from the sum of odd-indexed $a$-coefficients defined by the expansion of
the splittings on orthogonal polynomials
(e.g.  Schou et al. \cite{SCDT2})

\begin{equation}\label{eq:int}
\sum_{j=1,3,..}a_j^{nl}\simeq \int_0^{R_\odot} K_{nl}(r)\
 \Omega_{eq}(r)\ dr,
\end{equation} 
where $K_{nl}(r)$ are the so-called rotational kernels, which  have been
calculated for each mode from a solar model taken from Morel et al.
(\cite{morel97}). In the following, they are
 assumed to be known exactly.

This approximation of the 2D integral equation is valid only for 
high degree modes (e.g. Corbard \cite{corbard97:capo}) but the influence
of the low degree modes on the determination of the position and width of
 the tachocline and the rotation rate of the  upper layers is thought
to be small.

We discretize Eq.~(\ref{eq:int}) by using a polynomial expansion method,
which leads to the matrix equation: 

\begin{equation}
\vec W=\vec R\vec\Omega,
\end{equation}
where we have defined the vector 
$\vec W\equiv({W_i/\sigma_i})_{i=1,N}$,
of $N$ truncated sum of $a$-coefficients  $ W_i=\sum_{j=1}^{n_i} a_j$ 
weighted by the standard 
deviation $\sigma_i$ for each mode $i\equiv(n,l)$. 
The number $n_i$ of $a$-coefficients is fixed by the observations for 
each mode.
The standard deviations have been computed 
straightforwardly from the uncertainties quoted on each 
$a$-coefficient.
In this work,  we assume no error correlation between the different 
modes. 

We seek a solution $\bar\Omega(r)$ defined as a piecewise linear function
of the radius:
\begin{equation}\label{eq:expansion}
\bar\Omega(r)=\sum_{p=1}^{N_p} \omega_p\psi_p(r) \\ 
\vec\Omega\equiv (\omega_p)_{p=1,N_p}
\end{equation}
where $\psi_p(r)$, $p=1,N_p$ are piecewise straight lines ($N_p=100$ in this
work) between fixed break points distributed according to the
density of turning points of the modes (cf. Corbard et al. \cite{Corbard97}).
The discretization matrix $\vec R$ is then defined by: 
\begin{equation}\label{eq:discret}
\vec R\equiv (R_{ip})_{{i=1,N\atop p=1,N_p}} \\
R_{ip}={1\over \sigma_i}\int_0^{R_\odot} \!\!\!K_i(r)\psi_p(r) dr
\end{equation}

An  inverse method should lead to  a solution that 
is able to produce a good
fit to the data. We define the goodness-of-fit 
by  the $\chi^2$ value obtained for  any solution $\bar\Omega(r)$:

\begin{equation}
\chi^2(\bar\Omega(r))=
\sum_{i}\left[{W_i- \int_0^{R_\odot}K_i(r)
\bar\Omega(r)dr\over\sigma_i}\right]^2,
\end{equation}
which can be written in the discretized form:

\begin{equation}
\chi^2(\vec\Omega)=\|\vec R\vec\Omega-\vec W\|_2^2.
\end{equation}

\section{Regularization: the non linear approach}\label{sec:Reg_non_lin}
%
\subsection{generalized regularization term and Euler equations}
%
 Unfortunately, the  inverse integral problem is an ill-posed problem
and the minimization of only the $\chi^2$ value 
generally leads to
oscillatory solutions that are not `physically acceptable' in the sense that
they do not correspond to our a priori knowledge on  the shape of the solution.
 So, we have to
use regularization techniques i.e. to introduce a priori information in the 
minimization process. 
A large class of these techniques, can be expressed in the general form of
the  minimization of a criterion $J$ over the unknown solution
$\bar\Omega(r)$:

\begin{equation}\label{eq:crit}
J(\bar\Omega(r))=\chi^2(\bar\Omega(r))+\lambda^2\int_0^{R_\odot} 
\varphi\left({1\over\delta}{\left| d^q\bar\Omega(r)\over dr^q\right|}\right)dr,
\end{equation}
where $\lambda$ is the so-called trade-off parameter, 
chosen so that it  establishes  a balance between the goodness-of-fit to 
the data and the constraint introduced on the
solution. The parameter $\delta$
allows to fix the threshold on the gradient 
modulus of the solution under which it is smoothed, and above which 
it is preserved (cf. Sect.~\ref{sec:choix_reg_param}).
The order $q$ of the derivative is usually taken equal to one or
two. The two choices can lead to similar results
with the appropriate choice of the regularizing
parameter $\lambda$ in the domains where the solution is well constrained
 by the data. However these two choices correspond
to two different a priori constraints on the solution. As the
rotation is known to be quasi-rigid  in the radiative interior 
(at least down to $0.4R_\odot$),
we have chosen in this work  to use the first derivative. 
We note however that the method described below can easily be generalized
for any choice of the derivative order.

 Two choices for the
$\varphi$-function lead to well known regularization strategies: 
\begin{itemize}

\item
$\varphi(t)=t^2$ leads to the traditional Tikhonov approach with first 
derivative  whereas 
\item
$\varphi(t)=t$ is known as the \emph{Total Variation} (TV) regularization 
method
(e.g. Rudin et al. \cite{rudin92}; Acar \& Vogel \cite{acar94}). This 
approach  uses the absolute value rather than the square modulus 
of the solution gradient.  It has been shown that the solution is
 searched in a space composed of bounded variation  functions which admit 
discontinuity points. This regularization method is therefore   
able to recover piecewise smooth solutions with steep gradients 
(see e.g. Dobson \& Santosa \cite{dobson94}; Vogel \& Oman \cite{vogel96}, 
\cite{vogel97}). 
The PP-TSVD method of Hansen \& Mosegaard (\cite{PPTSVD}),
 already used in helioseismic
context in Corbard et~al. (\cite{corbard98}),
 can be seen as a `truncated version'
(in the sense that the regularizing parameter is a discrete
truncation parameter)
of the TV regularization  in the same way as the MTSVD
method introduced by Shibahashi \& Sekii  (\cite{MTSVD1})
 (see also: Hansen et al. \cite{MTSVD2}; Corbard et al. \cite{corbard98})
  is a  `truncated version'
of the Tikhonov regularization.

\end{itemize}

For a general $\varphi$-function, one can write  the
 criterion  Eq.~(\ref{eq:crit}) in a discretized form:

\begin{equation}
J(\vec\Omega)=\chi^2(\vec\Omega)+\lambda^2 J_2(\vec\Omega),
\end{equation}
where $J_2(\vec\Omega)$ represents the discretized
regularization term defined by:

\begin{equation}\label{eq:J2}
\int_0^{R_\odot}\!\!\!\varphi\left({1\over\delta}{\left| d\bar\Omega(r)\over 
dr\right|}\right)dr
= J_2(\vec\Omega)=\sum_{p=1}^{N_p-1} c_p 
\varphi\left({\left|\vec{L}\vec\Omega\right|_p\over \delta}\right).
\end{equation}
In this equation $(c_p)_{p=1,N_p-1}$ represent the weights used for the
integration rule, and $\vec L$ is a discrete approximation of the 
first derivative operator. $\left|\vec{L}\vec\Omega\right|_p$ is the  
absolute value of the $p^{{\rm th}}$ component of the vector $\vec{L\Omega}$.
 The expression of $c_p$ and $\vec L$
are given in Appendix  for the simple case of the polynomial
expansion Eq.~(\ref{eq:expansion}) 

The minimization of the criterion $J(\vec\Omega)$ over each component
$\omega_p$ of $\vec\Omega$ leads to the following Euler equations 
(discretized form):

\begin{equation}\label{eq:euler1}
 \nabla J(\vec\Omega)=0 \Longleftrightarrow
(\vec R^\top\vec R  +{\bar\lambda}^2 \vec L^\top \vec B(\vec \Omega) 
\vec L)\vec\Omega =
\vec R^\top \vec W 
\end{equation}
where $\bar\lambda={\lambda\over \delta}$ and $\vec B $ is 
a diagonal matrix with elements that  depend on the
gradient of the solution at each grid point: 

\begin{equation}\label{eq:euler2}
 \vec B=diag(b_p)\ \ \ \mbox{with}\ \ \ 
b_p=c_p {\varphi^{'}\left(t\right)\over 2 t} \ \ \ \mbox{and} \ \ \
t={\left|\vec{L}\vec\Omega\right|_p\over \delta}
\end{equation}

For $\varphi(t)=t^2$, $\vec{B}$ is independent of $\vec{\Omega}$ and 
these normal equations reduce to a 
linear system which corresponds to the usual Tikhonov regularization 
with first derivative.
On the other hand, for a general $\varphi$-function,
  this leads to a nonlinear problem which
requires an appropriate iterative method of solution. 

Now, with this general expression for the normal equations, the question
for our particular problem becomes: what properties must the  
$\varphi$-functions
satisfy to ensure  the preservation of  high gradients in the solution? 
The next section shows how the theoretical works developed in
the field of computed imaging gives an answer to this question,
and leads to an algorithm for solving the non linear Euler equation
 that is easy to implement.
\subsection{Properties of the weighting function ${\varphi^{'}(t)\over 2t}$}

From the Euler equation (Eqs. (\ref{eq:euler1}) and (\ref{eq:euler2}))
we can see that  the function ${\varphi^{'}(t)/2t}$ acts 
as a weighting function in the smoothing process: at each grid point
the gradient of the solution is used as an argument of this function
in order to set locally the magnitude of regularization. This suggests 
an iterative process where the gradient of the solution at a given 
step is used for the computation of the regularization term at the next step.
We show in the following that we can derive some basic properties 
of the  weighting function so that high gradients can be preserved,
and such that an iterative algorithm for solving the Euler equation is
 possible. First let us look at the behaviour of the weighting function
at the limits of low and high gradients:

\begin{itemize}

\item
For low gradients, we want to keep a Tikhonov regularization. From 
Eq. (\ref{eq:euler1}) this is the case if 
${\varphi^{'}(t)/ 2t}$ is a non-null constant function.

\item For high gradients, we want to remove smoothing.
This happens when ${\varphi^{'}(t)/ 2t}$ is close to zero.

\item 
Another property that sounds reasonable is to choose a decreasing function
of the gradient between these two limits. 
Furthermore, in order to avoid numerical
 instabilities
we will choose only strictly decreasing weighting functions.

\end{itemize}

Therefore the choice of the $\varphi$ function must be made taking into
account the following three properties needed on the weighting function
 ${\varphi^{'}(t)/2t}$ (Charbonnier et al. \cite{charbonnier94}, 
\cite{charbonnier97}):
\begin{enumerate}
\item Tikhonov smoothing for low gradients:

\begin{equation}\label{eq:ppty1}
0<\lim_{t\rightarrow0}{\varphi^{'}(t)\over 2t}=M<\infty 
\end{equation}

\item  no smoothing for high gradients:

\begin{equation}\label{eq:ppty2}
\lim_{t\rightarrow\infty}{\varphi^{'}(t)\over 2t}=0
\end{equation}

\item 

\begin{equation}\label{eq:ppty3}
\begin{array}{ll}   
{\varphi^{'}(t)\over 2t}  & \mbox{strictly 
decreasing on }[0,+\infty[\\
& \mbox{to avoid instabilities}.\nonumber 
\end{array}
\end{equation}

\end{enumerate}

Within these conditions, the $\varphi$-function may be chosen either
convex (Green \cite{green90}; Charbonnier et al. \cite{charbonnier94})
 or non-convex (Perona \& Malik \cite{perona90}; Geman \& McClure 
\cite{geman85}; Hebert \& Leahy \cite{hebert89}) 
(see also Charbonnier et al. (\cite{charbonnier97}) and Teboul et al. 
(\cite{teboul98}) for examples in both cases). A 
non-convex function may be better suited for the search of high gradients.
Nevertheless this choice leads to some numerical difficulties and instabilities
related to the existence of local minima. This may induce a high 
sensitivity of the inverse process to the choice of the regularization
parameters (Blanc-F\'eraud et al. \cite{blanc-feraud95}). 
On the other hand, the choice of a convex
function may avoid these numerical problems and is more suitable for relatively
smooth transition (Blanc-F\'eraud \cite{blanc-feraud98}).

We note that in the case of the TV regularization (or equivalently the
PP-TSVD method) the function $\varphi(t)=t$ does not satisfy the first 
property (Eq.~\ref{eq:ppty1}).
This may explain the difficulties encountered in using this
method (Corbard et al. \cite{corbard98}). 
For smooth transitions, the dispersion 
of the results for different realizations of the noise became large,
indicating some instabilities in the inversion process. In light of the above
formalism, this may be related to the non differentiability
 of the corresponding
weighting function $\varphi'(t)/2t$ at $t=0$, which can lead 
to numerical instabilities.

\subsection{The iterative algorithm: ARTUR}
\begin{figure}
  \resizebox{\hsize}{!}{\includegraphics[angle=-90]{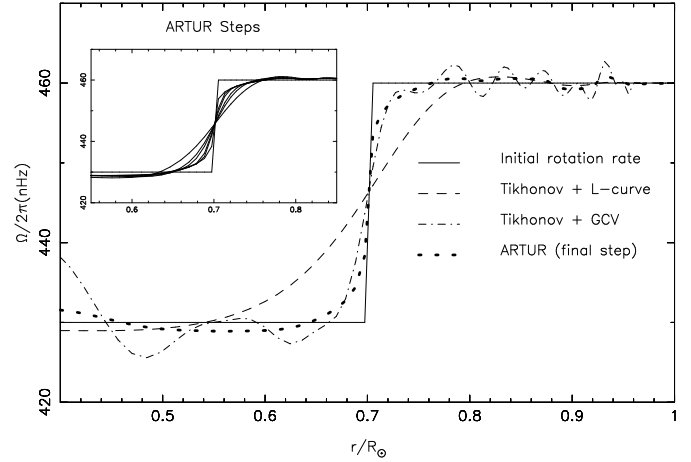}}
   \caption{Solutions obtained by inverting  splittings computed
from a discontinuous one dimension 
rotation profile (solid line) for the same modeset 
as in the LOWL data and including  some 'realistic' noise (see text).
The standard Tikhonov
solution is given for two different automatic choices of the regularizing 
parameter. 
The successive steps of ARTUR algorithm are shown in the upper left insert,
 whereas the final step is shown  on the main plot. 
The choice of the regularizing function and 
parameters for ARTUR  algorithm are those 
discussed  in Sect.~\ref{sec:choix_reg_param}. The solutions are plotted 
without  error bars for clarity. }
   \label{fig:comp.ks1}
\end{figure}

\begin{figure}
  \resizebox{\hsize}{!}{\includegraphics[angle=-90]{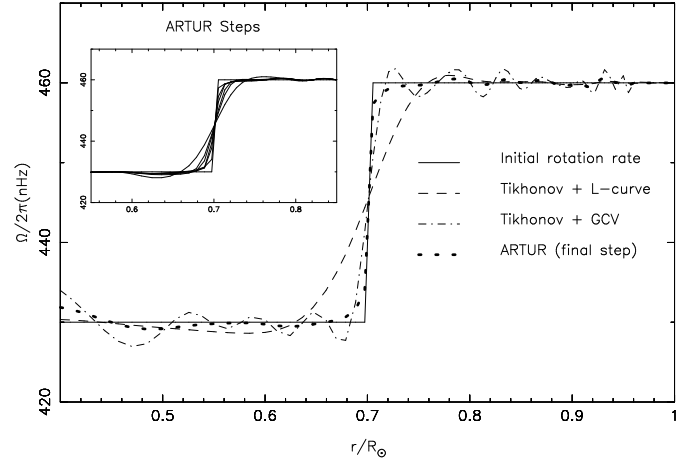}}
   \caption{The same as Fig.~\ref{fig:comp.ks1} but computed for a lower 
level of noise (standard deviations divided by $\sqrt{10}$). 
Comparison between Figs. \ref{fig:comp.ks1} and \ref{fig:comp.ks10} show 
the smoothing effect
of the data noise level for the three methods.}
   \label{fig:comp.ks10}
\end{figure}

%
Under the three conditions (\ref{eq:ppty1})(\ref{eq:ppty2})(\ref{eq:ppty3}),
 it has been shown
by Charbonnier et al. (\cite{charbonnier94}, \cite{charbonnier97}) 
that the non linear criterion can be solved by  using an iterative 
scheme named ARTUR (Algebraic Reconstruction Technique Using Regularization)
that is easy to implement:
at each step $k$ we calculate
the regularization term using the derivative of the previous 
estimate $\vec \Omega^{k-1}$ and  we simply
compute the new estimate $\vec \Omega^k$  by solving the linear system:
\begin{equation}\label{eq:lin_syst}
 \Bigl(\vec R^\top\vec R  +{\bar\lambda}^2 \vec L^\top 
\vec{B}(\vec \Omega^{k-1}) \vec L\Bigr)\vec\Omega^k =
\vec R^\top \vec W. 
\end{equation}
When we use a convex $\varphi$-function, the convergence of this so-called 
\emph{half quadratic algorithm} (minimization of a quadratic 
criterion at each step (Geman \& Reynolds \cite{geman92})) to the minimum of 
the criterion 
given by Eq. (\ref{eq:crit})
has been established (Charbonnier et al. \cite{charbonnier97}).
This is also an adaptive regularization method which uses the information
on the derivative of the solution obtained at each step in order
to improve the regularization at the next step. This requires an
initial guess for the solution, but we will show in the next section 
that a constant solution can always be used as the starting guess.
Figures~\ref{fig:comp.ks1} and \ref{fig:comp.ks10}  show examples of ARTUR
steps in the case of a discontinuous rotation rate and for two different 
levels of noise. At each step of the ARTUR algorithm the linear system
Eq.~(\ref{eq:lin_syst}) has been solved using an iterative conjugate gradient
method with Jacobi preconditioning (see e.g. Golub \& Van-Loan \cite{golub89};
Barrett et al. \cite{templates}) 
using $\vec\Omega^{k-1}$ as starting point. This leads to a very fast 
algorithm
where the number of conjugate gradient iterations needed to solve the 
linear system decreases at each ARTUR step.
The algorithm is stopped when the $2$-norm of the relative
difference between two solutions at two successive steps is  below $10^{-6}$
i.e.:
\begin{equation}\label{eq:stop}
{\|\vec\Omega^k-\vec\Omega^{k-1}\|_2\over\|\vec\Omega^k\|_2}\le  10^{-6}
\end{equation}


\section{The choice of $\varphi$-function and regularizing
 parameters for rotation inversion}\label{sec:choix_reg_param}

\subsection{The $\varphi$-function}

In the particular case of the determination  of the solar 
tachocline profile,
the uncertainty on the width of the transition zone is still large
(see Table 2 of Corbard et al. (\cite{corbard98}) for a summary of some
previous works). Therefore, according to the previous discussion,
we have chosen to consider a  convex regularizing 
$\varphi$-function, specifically   the one defined in 
Charbonnier et al. (\cite{charbonnier94}, \cite{charbonnier97}): 
\begin{equation}
\varphi(t)=2\sqrt{t^2+1}-2
\end{equation}
This function is close to
the absolute value function used in TV regularization and PP-TSVD method,
but, unlike the absolute value, has a quadratic behaviour near $0$ that
satisfies the requirement of 
Eq.~(\ref{eq:ppty1})
on the contrary of the absolute value. The weighting function, shown as 
dashed lines in Figs.~\ref{fig:delta} and \ref{fig:lowl_grad}, is given 
by:
\begin{equation}\label{eq:wf}
{\varphi^{'}(t)/ 2t}=\left(1+t^2\right)^{-{1\over 2}}.
\end{equation}

\subsection{The regularizing parameters}

The choice of the parameters $(\lambda,\delta)$ is an important point, as in
every regularization methods. For example, using a Generalized Cross 
Validation (GCV) strategy (Wahba \cite{GCV}) 
for the regularizing parameter of a Tikhonov 
model leads to an oscillating  solution while  using the L-curve
strategy produces a smoother solution. This choice influences evidently the
estimation of the width of the tachocline. It is intuitively  not
surprising as the model represents a priori information on the
solution (so the solution depends on this information).

Parameter estimation for regularizing problem is a
deep question and represents an active research area of its own
(eg. Lakshmanan \& Derin \cite{LAKS89}; Thompson et al. \cite{THOMP91};
Galatsanos \& Katsaggelos \cite{GALA91}). 
Some results on $(\lambda,\delta)$ parameters have been obtained for the
proposed model in the field of image processing 
(Chan \& Gray \cite{KCHA96}; Jalobeanu et al. \cite{RREST98}; Zerubia et al.
\cite{SPIE98}). However, it is 
still an open problem, and studies and results depend sensitively on the
considered inversion problem.

In the remainder of this
paper, a choice
of the parameters $(\lambda,\delta)$ is proposed, based on heuristic
considerations (the following two sub\--sec\-tions) and simulation results
(Sect.~\ref{sec:mc}). The 
application to real data is then discussed in Sect.~\ref{sec:real_data}.

\subsubsection{The choice of $\overline{\lambda}$: using automatic strategies}
If the initial guess is  a constant function, then according
to  property 1 (Eq.~\ref{eq:ppty1}) and Eq. (\ref{eq:wf}), 
$M=1$ and  the solution at
the first ARTUR step corresponds to a Tikhonov solution with 
$\overline{\lambda}$  as regularizing parameter.
It has been shown in Corbard et al. (\cite{corbard98}) that
the GCV strategy 
for the choice 
of the regularizing parameter in Tikhonov inversions leads
systematically to better results concerning the evaluation of
the tachocline parameters, as
 compared to the L-curve strategy (Hansen \cite{L-curve}).
In fact, the GCV strategy leads systematically to  less
smoothing than the L-curve 
approach ($\lambda_{Lcurve}\simeq 100*\lambda_{GCV}$
in that work) 
and therefore is more suited to the study of a region with high gradients.
Nevertheless, because of the low global 
regularization, this choice may lead to 
spurious oscillations below and above the tachocline 
(see Figs. \ref{fig:comp.ks1}, \ref{fig:comp.ks10}). The
ARTUR algorithm will tend to enhance the high gradients found
at the first step. Therefore it is important to start with a 
 solution smooth enough to  avoid spurious oscillations with high gradients.
The GCV choice for  $\overline{\lambda}$ may therefore not be well adapted for
ARTUR initial step, and some experiments have shown that, on the other hand,
the L-curve choice leads often to a solution which is too smooth
and does not allow the buildup of  the  high gradients expected during the
iterations. Nevertheless, the optimal choice of this parameter strongly
depends on the level of noise included in the data, and  therefore
 it is important to define an automatic  choice of this parameter 
so that we use the same strategy  for different
realizations of the noise in Monte-Carlo simulations 
(cf. Sect.~\ref{sec:mc}). 
The results of these simulations are shown in Fig.~\ref{fig:mc} with both
$\bar\lambda=\lambda_{Lcurve}/10$ and $\bar\lambda=\lambda_{GCV}$. The second 
choice has been retained for the application to  real data 
(cf. Sect.~\ref{sec:real_data}).

\subsubsection{The choice of $\delta$: using  a priori knowledge on
 the searched gradients}\label{sec:choix_delta}

\begin{figure}
  \resizebox{\hsize}{!}{\includegraphics[angle=-90]{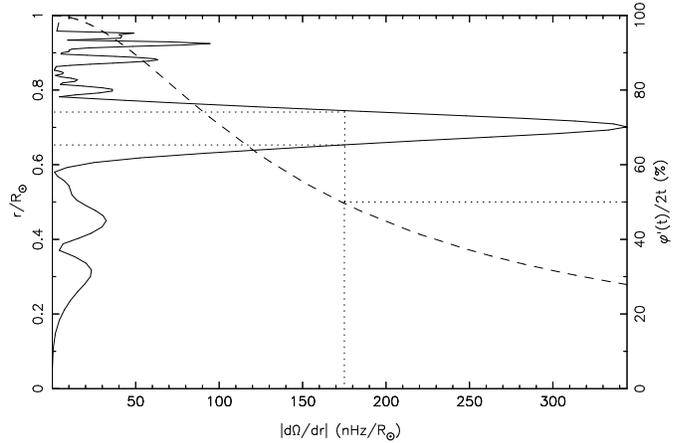}}
   \caption{The solid line shows the first derivative of a first step solution 
(cf. Fig.~\ref{fig:comp.ks1}) in the ARTUR algorithm 
as a function of the fractional solar radius. For each value 
of the gradient, the dashed line gives the weight that will be given locally 
to the regularizing term at the following step of the algorithm. The
weighting function is given by Eq.~(\ref{eq:wf}) with $t=|d\Omega/dr|/\delta$
and $\delta=100 R_\odot/$nHz. The dotted 
line indicates the gradient for which the local regularization will be 
reduced by  50$\%$  at the second step, as compared to the first step.
 Here, this  occur for radius
between $0.65R_\odot$ and $0.75R_\odot$.}
   \label{fig:delta}
\end{figure}				

The parameter $\delta$ has been introduced in order to adapt the shape
of the weighting function to the gradient that we seek to detect 
in a particular application. We have chosen for simplicity to keep
this parameter constant during the iterations. As we  start
the iterative process with a constant guess rotation, the first step is 
independent of $\delta$ and thus we can use the solution obtained 
after this  initial 
step in order to adapt the parameter $\delta$ to the gradients found 
in the first step solution. As we have shown that the first step of ARTUR 
algorithm correspond
to a classic Tikhonov inversion, this  choice of $\delta$ can be
viewed  as a way to use our a priori knowledge (as given by Tikhonov inversion)
of the searched gradients.

It is generally admitted that the width of the
tachocline does not exceed $0.1$ solar radius, which is also
the resolution typically reached near the tachocline localization 
 ($\simeq 0.69R_\odot$) with a Tikhonov method applied
to the current datasets.
  Furthermore we
have a good estimate  of the difference between the rotation rate
above and below the transition 
($\simeq 30$nHz in Corbard et al. (\cite{corbard98})). Therefore
we can estimate a level of  $300$nHz/$R_\odot$ 
for the maximum gradient obtained at the first iteration
of ARTUR process.

Figure \ref{fig:delta} shows (solid line)
the first derivative of solution obtained  at the first step
by inverting artificial splittings which have been computed for a 
discontinuous 
rotation law (cf. Fig.~\ref{fig:comp.ks1}) with a step of $30$nHz
and by adding some Gaussian noise with a standard deviation taken  
from the formal error given in LOWL data for each 
mode (cf. Corbard et al. \cite{corbard98}). The weighting function 
Eq.~(\ref{eq:wf})
is shown in dashed line for $\delta=100 R_\odot/nHz$.    
At the second step we want to preserve
only high gradients i.e.  to regularize less in these zones
where high gradients have already been found at the first
step. For example, according to Fig.~\ref{fig:delta}, 
the choice of $\delta=100 R_\odot/nHz$  leads
to regularize $50\%$ less
at the second step in that zones where the gradient of the first
step solution is above $\sim 175$ nHz/$R_\odot$. 
For the particular realization of the noise introduced in artificial data    
this choice of $\delta=100 R_\odot/nHz$ 
looks reasonable, in the sense that it will tend to decrease 
the regularization 
especially in the transition zone.
A smaller value would enhance the secondary peaks that are induced 
by the data noise. 

The maximum gradient obtained, at the first step, with the artificial dataset 
($350$nHz$/R_\odot$ in Fig.~\ref{fig:delta}) corresponds approximately to our
previous estimate of $300$nHz$/R_\odot$  expected for real data.
 For the Monte-Carlo 
simulations done in order to estimate the errors (cf. Sect.~\ref{sec:mc})
the parameter $\delta$ has been fixed to $\delta=100 R_\odot/nHz$. 
Nevertheless, 
 the shape of the solution derivative 
 after the first step is a function of the dataset through
the intrinsic resolution of the modeset and the level of noise.  
Therefore, with real data the choice of $\delta$ will always be made
by looking at the derivative profile after the first step. We will
see however in Sect.~\ref{sec:real_data} that other indicators can help
in the choice of $\delta$, and that according to these indicators
 the choice $\delta=100 R_\odot/nHz$ seems
to be a good compromise for LOWL data also.

\section{On the error estimation on tachocline parameters using nonlinear 
methods}\label{sec:mc}

\begin{figure}
  \resizebox{\hsize}{!}{\includegraphics[angle=-90]{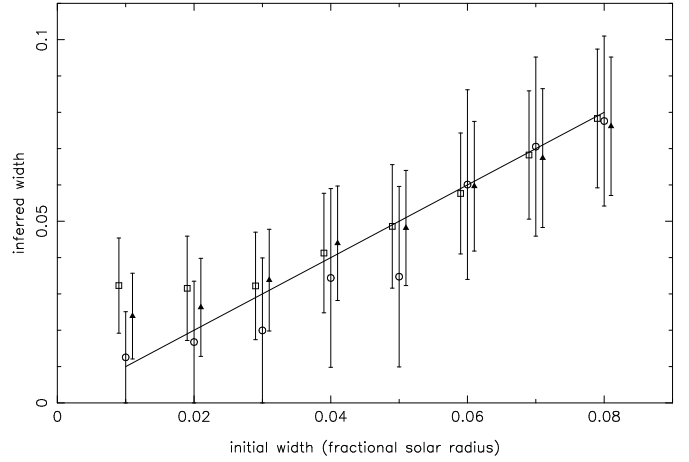}}
   \caption{Monte-carlo simulation for the estimation of the error on
the width inferred with the ARTUR algorithm. Triangles are for 
$\bar\lambda=\lambda_{Lcurve}/10$ and circles for 
$\bar\lambda=\lambda_{GCV}$. In both cases 
$\delta=100 R_\odot/nHz$. The rotation profile
was taken as an $erf$ function with different `initial widths'. The 
`inferred widths' are the mean value for 500 noise realizations  
of the results obtained by fitting 
directly the solutions
by an $erf$ function (cf. Eq.~\ref{eq:fiterf} where $\bar \alpha$ 
is assumed to be zero). Error bars 
represent a $68.3\%$ confidence interval on the width. For comparison, 
the squares show the result obtained from 
Tikhonov  method with GCV choice of the parameter after a `local deconvolution'
using the averaging kernels computed at the center of the transition 
(see Corbard et al. 1998). }
   \label{fig:mc}
\end{figure}

For nonlinear methods, we cannot compute straightforwardly the
formal errors at each point of the solution as we can do for linear
process. The ARTUR algorithm solves a  linear system at each step,
 but the final result depends nonlinearly on the data since
 the coefficients of the matrix to be inverted
at each step are functions of the data through the derivative of 
the previous estimate that is used as an argument of 
 the weighting function.

We focus here only on having
an estimate of the uncertainty on  the width $\bar w$ and the
 location $\bar r_c$ of the tachocline. 
We cannot obtain directly error
bars on the solution, and therefore the fit by  an $erf$ function does not
give an estimate of the error on the inferred parameters.
Instead, a first estimate of the uncertainty on 
the tachocline width inferred here has been computed using a Monte-Carlo
method applied on given rotation profiles simulated by $erf$ functions 
with widths lying
between $0.01R_\odot$ and $0.08R_\odot$ (in steps of $0.01R_\odot$)
and located at $ r_c=0.7R_\odot$.
The LOWL set of modes splittings corresponding to these rotation
 profiles have been computed  with addition of a Gaussian
 noise with a standard deviation taken for each 
mode from the formal error given in LOWL
data.
Since the
 ARTUR algorithm is non linear, we can not define averaging kernels but
Fig.~\ref{fig:mc} shows that the final step of the algorithm with 
$\bar\lambda=\lambda_{Lcurve}/10$ and $\delta=100 R_\odot/nHz$ leads 
directly to results
similar to those reached by Tikhonov inversion after a `local
deconvolution' using the averaging kernels computed at the center of 
the transition (see Corbard et al. \cite{corbard98}). 
The $1\sigma$ error interval on the width is found to be
around  $\pm 0.02R_\odot$ for widths in the range $0.01 - 0.08R_\odot$.
This uncertainty is   probably related to the intrinsic
resolution of the modeset  and the level of noise
contained in LOWL data.

A large bias is found with deconvolved Tikhonov method for
 $w \leq 0.02R_\odot$.
This bias is less important for the ARTUR method with $\bar\lambda=
\lambda_{Lcurve}/10$ and vanishes if we set  
$\bar\lambda=\lambda_{GCV}$. Nevertheless, in this latter case 
error bars  are bigger (around $\pm 0.025R_\odot$ in the range
 $0.02-0.08R_\odot$  of initial 
widths) and the inferred widths tend to be underestimated for
 initials widths between $0.03R_\odot$ and $0.05R_\odot$. 
This is due to the fact  that,  in mean  for the  simulated data, 
the GCV criterion leads to a too strongly oscillating solution
that is not a good starting point for ARTUR algorithm, as it tends to 
produce  very sharp 
transitions in a large number of realizations for all initial widths 
below $0.05R_\odot$.

The Monte-Carlo
simulations have been performed for $500$ realizations of the input errors
 for each initial width and the mean value over all the realizations 
($8\times500$) of the inferred center $\bar r_c$ is $0.701\pm 0.004R_\odot$.
The center of the $erf$ function seems  therefore to be very well recovered by
the inversion. Nevertheless, we must keep in mind
that the center of the tachocline is defined as
the center of the $erf$ function that gives the best fit of  the solution. 
This may not give an  appropriate view of the tachocline profile
if, for example,  
the solution is  found
to lack such a  symmetry in the lower and upper parts of the tachocline when 
inverting 
real data.

Another important point  that  
 Fig.~\ref{fig:mc} demonstrates is the ability of the ARTUR algorithm to 
recover not only rotation with a discontinuity (as shown in the 
examples of Fig.~\ref{fig:comp.ks1} and \ref{fig:comp.ks10}), 
but also rotation with
a relatively smooth transition. This property does not characterize the 
PP-TSVD method, and this was the reason for the difficulties encountered
by Corbard et al. (1998)
in interpreting the results obtained with this first non linear approach.

 Even if both methods give similar results 
in the mean for some choices of the regularizing parameters,
 it sometimes happen that the  two solutions differ strongly 
for a particular realization
of the noise. Furthermore, the two approaches are very different in their
underlying principles and therefore it is very interesting to compare the two 
results
with observed data. 

\section{Application to observed data}
\label{sec:real_data}
\begin{figure}
  \resizebox{\hsize}{!}{\includegraphics[angle=-90]{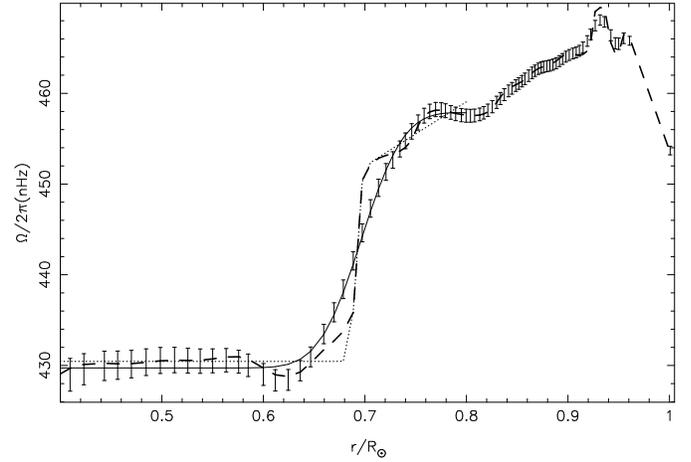}}
   \caption{Solar equatorial rotation rate estimated from LOWL data.
The vertical error bars  given 
at each grid points are the $1\sigma$ confidence intervals estimated for
the T-GCV solution. The dashed line shows ARTUR solution obtained 
with $\bar\lambda=\lambda_{GCV}$ 
and $\delta=100 R_\odot/nHz$. The fits by an $erf$ function 
(Eq.~(\ref{eq:fiterf}))
 of these two solutions are shown respectively by the solid and  dotted
lines. The fits have been computed
 only between $0.4R_\odot$ and $0.8R_\odot$ and the 
five parameters deduced from each fit are shown in 
Fig.~\ref{fig:lowl_delta_2} }
   \label{fig:lowl_100}
\end{figure}

\subsection{LOWL data}
%
The LOWL instrument is a Doppler imager based on a Potassium 
Magneto-Optical Filter that has been operating on Mauna Loa, Hawaii 
since 1994 (see Tomczyk et al. (\cite{tomczyk95}) for a detailed description).
The dataset includes the frequency splittings of $1102$ 
modes $(n,l)$ with degrees up to $l=99$ and 
frequencies lower than $\nu=3500\ \rm{\mu Hz}$ deduced 
 from a two year period of 
observation ( 2/26/94 - 2/25/96 ).
For each mode, individual splittings are given by, at best, 
$n_i=5$ $a$-coefficients of the expansion on orthogonal polynomials 
defined by Schou et al. (\cite{SCDT2}). 

\subsection{The choice of inversion parameters 
             and ARTUR solution for LOWL data}
%
 \begin{table*}
\caption[]{Comparison between our previous work on inferring the equatorial 
tachocline profile from LOWL data and the results obtained with non linear 
regularization applied to the same data. The previous estimates of the 
parameters and their errors have been
deduced from a comparison of three inversion methods, including the T-GCV 
method. See Sect.~\ref{sec:w} for
 a more detailed discussion on the tachocline
width using the ARTUR algorithm}
\begin{flushleft}
\begin{tabular}{llllll}
\hline\noalign{\smallskip}
& $\bar\Omega_0$(nHz) & $\bar\Omega_1$(nHz) & 
$\bar\alpha$(nHz$/R_\odot)$ & $\bar{r_c}/R_\odot$ & $\bar{w}/R_\odot$ \\
\noalign{\smallskip}
\hline\noalign{\smallskip}
Corbard et al. (1998) & $431.0\pm 3.5$ & $459.0\pm 1.5$ & 
$0$ & $0.695\pm 0.005$ & $0.05\ \pm 0.03 $\\ 
ARTUR  & 430.5 & 452.0 & 70.0 & $0.691$ & $0.01$\\
\noalign{\smallskip}  
\hline
\end{tabular}
\end{flushleft}
\label{tab:res}
\end{table*} 

Since the first step of the ARTUR algorithm is standard Tikhonov inversion,
we first present 
the results obtained from Tikhonov method and GCV choice of the
regularization parameter (called T-GCV method hereafter). 
The equatorial profile 
 obtained from T-GCV method on LOWL data  
is shown in Fig.~\ref{fig:lowl_100}. 
The fit of the solution with an $erf$ function of the form 
Eq.~(\ref{eq:fiterf}) leads to a width of $w\simeq 0.09R_\odot$
(cf. Fig.~\ref{fig:lowl_delta_2}d).
 By taking into account the width of the averaging
kernel computed at $0.7R_\odot$, the corrected inferred width
obtained from this `local deconvolution' is $w\simeq 0.06R_\odot$.

Contrary 
to the example shown in Fig.~\ref{fig:comp.ks1} for simulated data,
the GCV choice of the regularization parameter does not lead, with LOWL data,
to an oscillating solution. This may indicate that this particular realization
of the noise introduced in simulated data is rather different 
from the noise contained in  LOWL data. The formal errors
quoted on each $a$-coefficient are perhaps overestimated. Furthermore
 our model 
assumes that the errors are uncorrelated, which is probably not strictly the
case. Therefore, with real data, the T-GCV solution may be a good starting 
point for the ARTUR algorithm and we choose to take $\bar\lambda=\lambda_{GCV}$
so that the first ARTUR iteration leads to the T-GCV solution. 

\begin{figure}
  \resizebox{\hsize}{!}{\includegraphics[angle=-90]{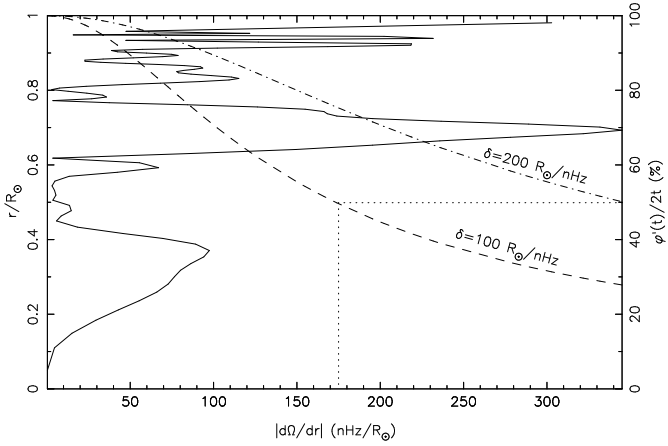}}
   \caption{Same as Fig.~\ref{fig:delta} but for LOWL data. The weighting 
function is shown for two choices of the parameter $\delta$. }
   \label{fig:lowl_grad}
\end{figure}

Figure~\ref{fig:lowl_grad}
shows the absolute value of the first derivative of the T-GCV solution. 
High gradients are found not only in the tachocline, but also
 near the surface (above $0.09R_\odot$). 
However
 LOWL data include relatively few high degree modes ($l\le 99$ in these data),
so  we will focus only on the tachocline in this work.   
As  discussed in Sect.~\ref{sec:choix_delta}, Fig.~\ref{fig:lowl_grad}
 can help for the choice of the parameter
$\delta$. The weighting function is shown for two values of $\delta$.
The choice $\delta=100 R_\odot/nHz$ lowers  the regularization
 by more than a factor two  at the second step
in the tachocline and in the upper layers,
 whereas the choice $\delta=200 R_\odot/nHz$
will never decrease the regularization by more than $50\%$ after the 
first step. From this figure, we can guess that a choice of 
$\delta<100 R_\odot/nHz$ will 
tend to enhance spurious oscillations due
to the noise, whereas a choice   $\delta>200 R_\odot/nHz$ will lead to a 
result very similar
to the T-GCV solution because only few points will be affected by the 
local change of regularization during the ARTUR steps. Between these two limits
however, it is not clear which is the best choice for $\delta$. 

\begin{figure}\LabelFig{a-c}{fig:lowl_delta_1}
  \resizebox{\hsize}{!}{\includegraphics{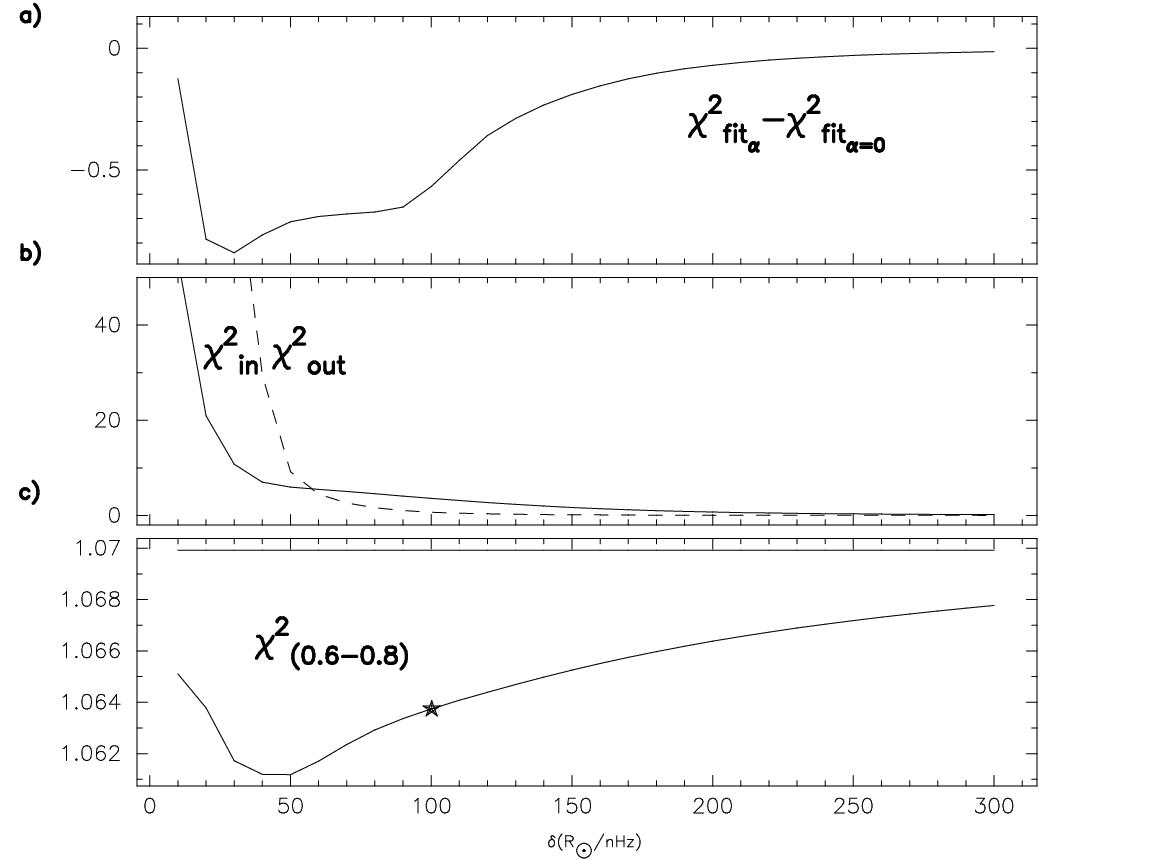}}
   \caption{Variation of various $\chi^2$ indicators with 
$\delta$. {\bf a}~Difference between the goodness of the fits of ARTUR
 solutions by an $erf$ function with or without a linear part 
(cf. Eq.~(\ref{eq:chi2fit})). {\bf b}~Difference between T-GCV and 
ARTUR solutions in and out of the zones where
high gradients are expected (cf. Eq.~(\ref{eq:chi2_in_out})). {\bf c}~The 
normalized $\chi^2$ value of ARTUR inversions for modes which have their
turning points between $0.6R_\odot$ and $0.8R_\odot$. The horizontal line
shows the value ($1.07$) obtained from T-GCV solution. The star symbol 
indicate 
the $\chi^2$ value reached for the choice $\delta=100 R_\odot/nHz$ retained 
for the ARTUR 
algorithm.}
\end{figure}

However, some indicators may help in the choice of $\delta$:
\begin{itemize}
\item First, an important test for any global inversion is its capability of
providing a good fit to the data.
Figure~\ref{fig:lowl_delta_1}c shows, as a function of $\delta$, 
the normalized $\chi^2$ value for  the modes which have their
turning points between $0.6R_\odot$ and $0.8R_\odot$. As expected, this value
is always lower for all ARTUR solutions than for the T-GCV solution, because
we tend to regularize less. The gain in the $\chi^2$ value
is however small because the regularization is  decreased only locally in 
a small region.
The $\chi^2$ value reaches a minimum for $\delta=50 R_\odot/nHz$ but, as 
expected from 
Fig~\ref{fig:lowl_grad}, for such 
low $\delta$, the ARTUR solution becomes very 
oscillating and a discontinuity ($w\simeq 0$) is found near the tachocline but
in fact the solution is found to be piecewise constant with many 
discontinuities.

\item There is another indicator that we can use showing that this value
of $\delta=50 R_\odot/nHz$ is not appropriate, despite the good $\chi^2$ 
value reached
with this parameter. One of the objectives of the 
ARTUR method is to keep the same 
regularization as in Tikhonov method  in zones without high  gradients.
  Therefore one expects 
that the amount of change (compared to T-GCV solution)
 will be more important in the tachocline (and
possibly near the surface) than in other zones. 
We denote by  
$\bar\Omega_A^\delta(r)$ the solution at the final step of ARTUR algorithm 
for a given  $\delta$, and by $\bar\Omega_T(r)$ the T-GCV solution; we then
define 
the two quantities:
\begin{equation}\label{eq:chi2_in_out}
\chi^2_{\left|{in\ \atop out}\right.}=
{1\over N_{\left|{in\ \atop out}\right.}}\sum_{p\in I_{\left|{in\ \atop out}
\right.}}\left(\bar\Omega_A^\delta(r_p)
-\bar\Omega_T(r_p)\right)^2
\end{equation}
where $I_{in}=\{p\ /\ 0.6\!<\!{r_p\over R_\odot}\!\le \!0.8\  
\mbox{or}\ r_p\!>\!0.9R_\odot\}$, $I_{out}=[1,N_p]-I_{in}$  and $N_{in}$, 
$N_{out}$ are the 
sizes of the two sets.
Figure~\ref{fig:lowl_delta_1}b
 is a plot of these two  quantities. It shows  that 
for $\delta$ lower than $60 R_\odot/nHz$ the ARTUR 
algorithm tends to alter the solution even in regions
where no high gradients are expected,  but that for values above 
$\delta\simeq 100 R_\odot/nHz$ there is no
longer any changes in these regions, as expected.
\end{itemize}

From these two criteria it seems
that a choice of $\delta=100 R_\odot/nHz$ is a good compromise between 
minimizing the 
$\chi^2$ value for modes which have their turning points near the tachocline,
and operating changes essentially in zones where high gradients
are expected. This was precisely the objective of the non linear 
regularization 
approach. The ARTUR solution is shown in Fig.~\ref{fig:lowl_100} for this 
choice $\delta$. The inferred tachocline parameters are summarize in 
Tab.~\ref{tab:res} and compared to the results obtained by 
Corbard et al. (1998) for
the same dataset.  The estimates of the center of the tachocline and 
the rotation rate in the radiative interior 
obtained from ARTUR algorithm 
are in good agreement with our previous work. The inferred value of 
$\bar\Omega_1$ and 
$\bar\alpha$  are such that 
$\bar\Omega_1+0.1\bar\alpha\simeq 459$nHz, which correspond
to the rotation rate inferred at $0.8R_\odot$ and to the value of the
$\bar\Omega_1$ parameter obtained by the fit of the T-GCV solution. 
The tachocline width inferred from the ARTUR inversion  ($w=0.01R_\odot$)
is 
smaller than our previous estimate but still compatible if we take into 
account
an error of $0.02R_\odot$ as indicated by the Monte-Carlo simulations.
The next section will give a more detailed discussion
for the interpretation of  this result by showing 
 how the estimates of the width vary with 
the $\delta$ parameter, and during the  ARTUR iterations.

\subsection{On the inferred tachocline width}\label{sec:w}
%
\begin{figure} \LabelFig{a-d}{fig:lowl_delta_2}
  \resizebox{\hsize}{9cm}{\includegraphics{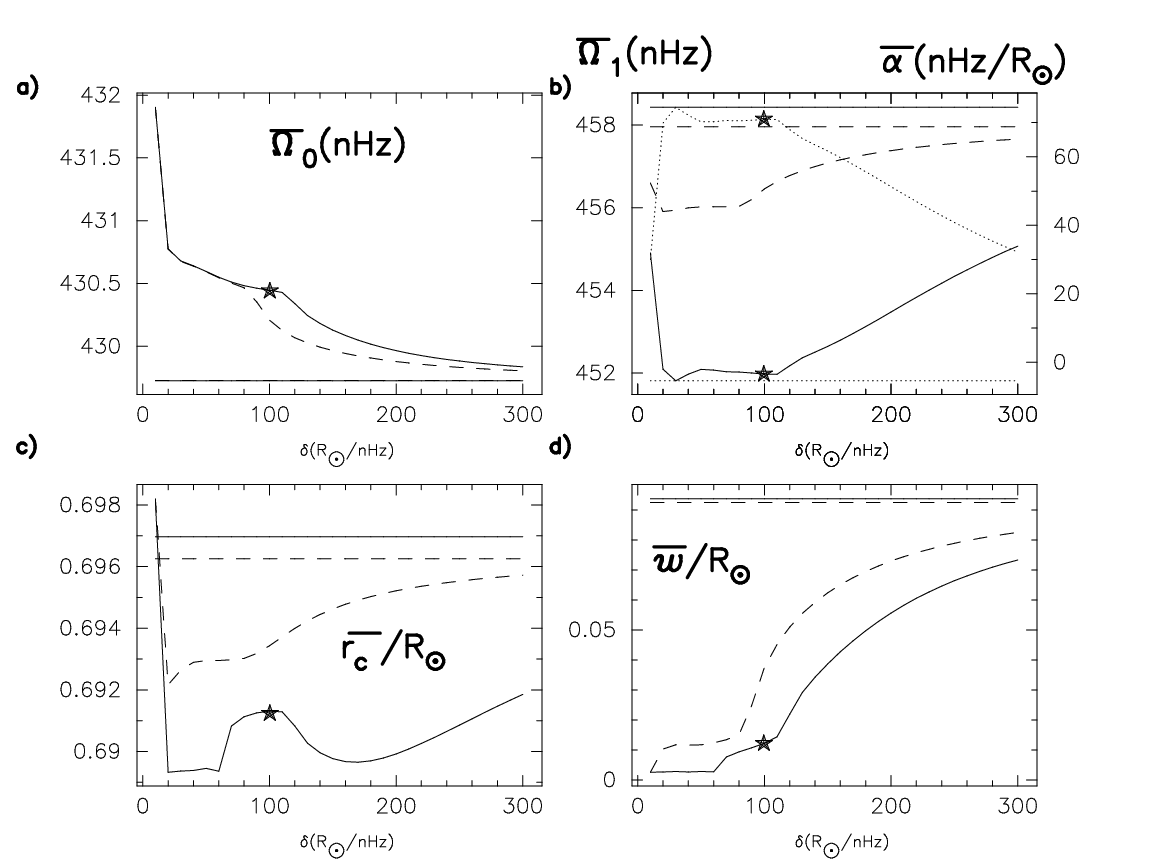}}
   \caption{Variation with $\delta$ 
of the tachocline parameters as deduced by 
fitting ARTUR solutions by an $erf$ function (cf. Eq.~(\ref{eq:fiterf})).
 The dashed lines show the results
obtained by searching only four parameters ($\bar\Omega_0$, $\bar\Omega_1$, $\bar r_c$,
$\bar w$) after setting $\bar \alpha=0$, whereas the solid lines show 
the results obtained 
when $\bar \alpha$ is a free parameter of the fit. In this latter case,
 the inferred value
of $\bar \alpha$ is shown by the dotted line on panel b and the star symbols 
show the results
obtained for $\delta=100 R_\odot/nHz$. The horizontal lines 
indicate the values (independent of $\delta$)
 of the tachocline parameters obtained by fitting the T-GCV
solution. These horizontal lines represent limits for high $\delta$ of 
the parameters
inferred from ARTUR solutions. On panel d the horizontal lines indicate
the width as inferred directly by the fit of T-GCV solution.
 The width corrected by a
`local deconvolution' using averaging kernels, is $w\simeq 0.06R_\odot$.  }
\end{figure}

\begin{figure} \LabelFig{}{fig:lowl_iter}
  \resizebox{\hsize}{!}{\includegraphics[angle=-90]{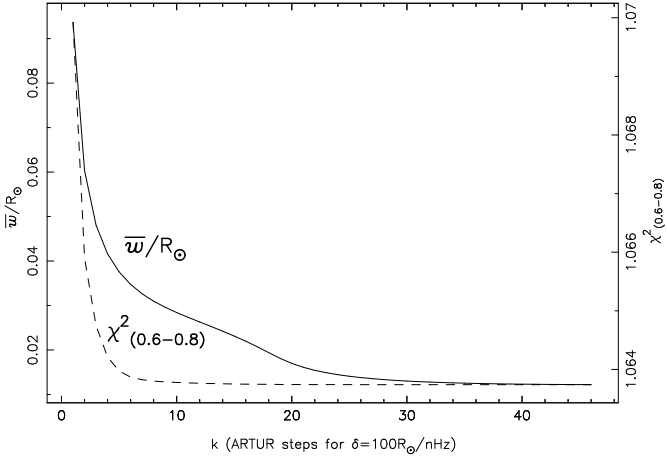}}
   \caption{Variation of the inferred width (solid line) and the $\chi^2$ 
value for modes which have their turning points between 
$0.6R_\odot$ and $0.8R_\odot$ (dashed line) as a function of the iteration 
number in ARTUR process for $\delta=100 R_\odot/nHz$). The values found at the 
first step are equal to those obtained from the fit of the T-GCV solution 
shown as horizontal lines in Figs.~\ref{fig:lowl_delta_2}d and 
\ref{fig:lowl_delta_1}c whereas the values obtained at the final step are 
shown 
by star graph markers on these plots.}
\end{figure}

Figure~\ref{fig:lowl_delta_2} 
shows how the  parameters, inferred from a fit of the final step
of the ARTUR algorithm are sensitive to the choice of $\delta$. It shows 
that the relation $\bar\Omega_1+0.1\bar\alpha\simeq 459nHz$ is 
still valid for other choices of $\delta$, and that $\bar\Omega_0$ and 
$\bar r_c$
vary only little with $\delta$ whereas the inferred width increases 
rapidly for $\delta>100 R_\odot/nHz$. It reaches $0.055R_\odot$
for $\delta=200 R_\odot/nHz$ which is the limit for the choice of $\delta$ 
above which
we think that the ARTUR algorithm is no longer effective.  
For large values of $\delta$, the number of step in the ARTUR process becomes 
very low, ARTUR solutions
tend to T-GCV solution, and thus, 
tachocline parameters inferred from ARTUR 
solutions tend toward those inferred from T-GCV solution 
(cf. Fig.~\ref{fig:lowl_delta_2}).

We have tried 
to fit the solutions with or without the linear part after the transition 
(i.e. by searching for the best $\bar\alpha$ coefficient or by setting 
$\bar\alpha=0$ 
in Eq.~(\ref{eq:fiterf})).
 The goodness of the fit 
is defined by:
\begin{equation}\label{eq:chi2fit}
\chi^2_{fit_{\bar\alpha}}(\bar\Omega)=
{1\over N_{fit}-5}\sum_{p\in I_{fit}}
\left(\bar\Omega(r_p)-\Omega_{fit}(r_p)\right)^2
\end{equation}
where $I_{fit}$ is the set of indices given by
 $I_{fit}\equiv\{p\ /\ 0.4\le {r_p\over R_\odot}<0.8\}$, and $N_{fit}$ is the 
size of
 $I_{fit}$. We denote by
$\chi^2_{fit_{\bar\alpha=0}}$ the goodness of a fit obtained
with only four parameters
($\bar\alpha=0$ in Eq.~(\ref{eq:fiterf})). In this case
 the denominator of Eq.~(\ref{eq:chi2fit}) becomes
$N_{fit}-4$.
The two fits are almost equivalent when applied to the T-GCV solution 
($\bar\alpha=-5$nHz when it is searched, cf. Fig.~\ref{fig:lowl_delta_2}b),
 but the fit that allows
a linear part has been found to be better suited for describing 
ARTUR solutions for all the choices of $\delta$ 
(cf. Fig.~\ref{fig:lowl_delta_1}a). We note however that if 
one chooses to fit the solutions 
 with $\bar\alpha=0$ (i.e. to describe the tachocline by a simple $erf$ 
function), then the inferred width would be systematically 
increased by a value up to  $0.02R_\odot$ 
(cf. Fig.~\ref{fig:lowl_delta_2}d).

 As the increase of the 
$\chi^2$ value between $\delta=100 R_\odot/nHz$ and $\delta=200 R_\odot/nHz$ 
is low 
(cf. Fig~\ref{fig:lowl_delta_1}a), and because of our previous estimate
of $\pm 0.02R_\odot$ for the uncertainty on the width,
we cannot exclude a width larger than $0.01R_\odot$. 
Furthermore Fig.~\ref{fig:lowl_iter} shows that the minimum value of the
$\chi^2$ is reached after only six ARTUR iterations, whereas 
the inferred width still decreases from $0.035R_\odot$ down to
$0.01R_\odot$. This indicates that the data themselves do not
allow us to choose between widths in that range. During the iterations
as well as when we vary the value of $\delta$, it is the amount 
of regularization introduced in high gradients zones
that is changed. In that sense, stopping
ARTUR iteration before its  convergence
according to the criterion Eq.~(\ref{eq:stop}) would be equivalent to increase
the value of $\delta$. As the needed amount of regularization 
is related to the level of noise contained in the data, the reliability
of our result is strongly related to  our
knowledge  of the data noise. Since we have
shown that the GCV criterion may reveal some discrepancies between
the data noise and the simulated noise, the only way to gain  more 
confidence on the appropriate choice of $\delta$, and thus on the result
concerning the tachocline width inferred from this kind of non linear 
inversions,
 will be  to increase our knowledge of the statistical properties of the 
data noise and to compare with other datasets.   
However, even the values of the width found at the final step
for $\delta=200 R_\odot/nHz$ or at the
sixth iteration with $\delta=100 R_\odot/nHz$ are still
below the value of $0.06R_\odot$ inferred from T-GCV solution after
a `local deconvolution'.  The use of nonlinear regularization argue
in favor of a very sharp tachocline and even 
a discontinuity can not be excluded.
Therefore our conclusion on the tachocline width
is  that it is very likely that it 
is less than $0.05R_\odot$. This is reinforce by the simulations (circles 
in Fig.~\ref{fig:mc}) showing that  initial widths up to $0.05R_\odot$
can lead to inferred widths of $0.01R_\odot$ whereas it is excluded 
(within $1\sigma$ error bars) for initial widths larger than $0.05R_\odot$.

\section{Conclusions}\label{sec:conclu}

This work introduces  in
helioseismic context an  approach of the inverse problem that use 
an adaptive regularization which is then used toward on the estimation of the
width of the tachocline. We do not claim that this paper gives a
definitive answer to this question. Instead, this paper presents a new tool
which allows to accurately reconstruct rotation profiles 
 with possibly both smooth and abrupt 
variations with depth.  This approach leads to a non linear 
problem that can be solved easily by an iterative process named ARTUR,
initially developed in the field of image processing.
It is shown that this allows to recover
high gradients in the solution and avoids the spurious oscillations
(known as `Gibbs phenomena') that may be found when we try to recover
such sharp transition zones with the usual Tikhonov approach.

The   proposed procedure for choosing the 
regularizing parameters and the Monte-Carlo simulations
 represent  a first
step showing the feasibility and the capability of 
the method to retrieve both small and large tachocline 
widths from noisy observations. They
have shown that this method, as well as the Tikhonov inversion
with `local deconvolution'  is able to recover the width of 
$erf$ functions with widths from  to $0.03R_\odot$ up to
 $0.08R_\odot$ with the same
error estimation of $\pm 0.02R_\odot$. For lower values of the
 width, the results are more sensitive to the choice of the
 regularizing parameters.  However, some improvements may be
 brought by the  studies on the optimal
choice of the regularizing parameters that are underway 
and their future application to the rotation
inverse problem.

The inversion of LOWL data using the  ARTUR method gives an equatorial 
tachocline profile which differs from our previous work.
 These new results  favor a
sharp transition (down to  a width of $0.01R_\odot$ with
the adapted regularization parameters). 
 From our study of the inferred width as a function of the
parameter $\delta$
and from our estimate of the uncertainty on this parameter, we conclude
 that the 
width of the tachocline should be less than $0.05R_\odot$. This estimate 
is somewhat
different from  our previous estimate of $0.05\pm 0.03R_\odot$ which allows
relatively smooth transitions up to $0.08R_\odot$. 

 The change in our estimate of the width is partly due 
to the fact that we have changed the fitting function that defined
the tachocline parameters. It is shown here that adding a linear
behaviour in the upper part of the $erf$ function allows a better fit 
of the solution.  Whereas the T-GCV solution (before any deconvolution) 
can be well approximated by a simple
$erf$ function between $0.4R_\odot$ and $0.8R_\odot$, the ARTUR
solution is better approximated, in the upper part of the tachocline, 
by a linear function with a slope around $70$nHz$/R_\odot$ that goes from 
$452$nHz at $0.7R_\odot$ up to $459$nHz at $0.8R_\odot$. 
It was not possible to reach this conclusion
from our previous approach using the T-GCV solution because
the `local deconvolution' used in this approach supposed
 explicitly that the rotation profile can be well approximated
by a simple $erf$ function (without linear behaviour).  
It will be
therefore very interesting to study in future works and with other datasets
the rotation profile just above the tachocline, in order to become more 
confident of our result.

An important  contribution of the non linear regularization
approach is that it allows to find directly a solution with sharp transitions
without fixing a priori the shape of the transitions, as  necessary
in forward modeling or when we deconvolve a solution obtained from linear
methods. In order
to describe the tachocline with only few parameters, we can afterwards 
choose a shape for the fitting function that is adapted to the solution
found.

Finally, we note that this  approach to inversion with non linear
regularization may find other applications in the helioseismic context,
for any the problems where high gradients are expected in the solutions.
We have shown that this is the case for the rotation of the surface layers,
which can be studied more accurately with instruments such as MDI on 
board SoHO,
or the GONG network which observe high degree modes. This algorithm
may also be applied
 to the sound speed anomaly found between the real Sun and solar models 
by structural inversions. The width of this peak, which is 
 located in the tachocline,
can be related to the  width of the mixed zone which is supposed 
to exist just below the convection zone (e.g. Morel et al. 
\cite{morel_boston98}). Some recent work (Elliott et al. 
\cite{elliott_boston98}) have 
used a linear inversion which has been deconvolved to give
an estimate of $0.02R_\odot$ for this width.  It is interesting
to note that this is of  the same order as 
our estimate of the tachocline width deduced from the rotation profile.
As  the sound speed anomaly 
profile is also a zone with high gradients in the solution of an inverse
problem, the use of 
non linear regularization may also
be an alternate approach to address this problem in future work.

\begin{acknowledgements}
%
We gratefully thank S.~Tomczyk and J.~Schou for providing
the LOWL data, and the referee for constructive remarks.
This work has been performed using the computing facilities provided 
by the program
``Simulations Interactives et Visualisation en Astronomie et M\'ecanique''
(SIVAM, OCA, Nice) and by the ``Institut du D\'eveloppement 
et des Ressources en Informatique Scientifique'' (IDRIS, Orsay).

\end{acknowledgements}

\appendix
\sectionApp{ terms of the discretization}
In our application, the polynomial
expansion Eq.~(\ref{eq:expansion})
 is such that:
\begin{equation}\label{eq:splines}
\psi_p(r)=\cases{ {r-r_{p-1}\over r_p-r_{p-1}} & if $r_{p-1}<r\le r_p$ \cr
                 {r_{p+1}-r\over r_{p+1}-r_p} & if $r_{p}<r\le r_{p+1}$\cr
0 & otherwise}
\end{equation}
where $(r_p)_{p=0,N_p+1}$ are the fixed break points 
distributed according to the
density of turning points of modes. We have: 
\begin{equation}
0=r_0=r_1<r_2<..r_{N_p-1}<r_{N_p}=r_{N_p+1}=R_\odot,
\end{equation}
therefore, each coefficient $\omega_p$ of the expansion 
Eq.~(\ref{eq:expansion})
simply represents the solution at the radius $r_p$:
\begin{equation}
 \forall\ p=1,..N_p \ \ \bar\Omega(r_p)=\omega_p. 
\end{equation}
Furthermore,
 with  this expansion of the solution, the first derivative 
of $\bar\Omega(r)$ is represented by a piecewise constant function.
Therefore, in this trivial case, one can take in Eq.~(\ref{eq:J2}):
\begin{equation}
c_p=r_{p+1}-r_p,\quad p=1,..N_p-1 \qquad \vec{C}\equiv diag(c_p) 
\end{equation}
and  the first derivative operator is defined by the bi-diagonal matrix:

\begin{equation}
\vec L=\vec{C}^{-1}\vec{L}^{(1)}
\end{equation}
with:
\begin{equation} 
\vec{L}^{(1)}_{i,j}=-\delta_{i,j}+\delta_{i,j-1} \left\{
\begin{array}{l}
i=1,N_p-1 \\
j=1,N_p
\end{array}\right. ,
\end{equation}


\end{document}